\documentclass[article,preprints,pdftex,moreauthors]{Definitions/mdpi} 

\firstpage{1} 
\pubvolume{1}
\issuenum{1}
\articlenumber{0}
\pubyear{2024}
\copyrightyear{2024}
\datereceived{ } 
\daterevised{ } 
\dateaccepted{ } 
\datepublished{ } 
\hreflink{https://doi.org/} 

\Title{Parameter optimization of Josephson parametric amplifiers using a heuristic search algorithm for axion haloscope search}

\TitleCitation{Characterization of Josephson parametric amplifiers using a heuristic search algorithm for axion haloscope search}

\Author{Younggeun Kim$^{1\dag}$, Junu Jeong$^{1\dag}$, SungWoo Youn$^{1*}$, Sungjae Bae$^{2,1}$, Arjan F. van Loo$^{3,4}$, Yasunobu Nakamura$^{3,4}$, Sergey Uchaikin$^{1}$, Yannis K. Semertzidis$^{1,2}$}

\address{%
$^{1}$ \quad Center for Axion and Precision Physics Research, IBS, Daejeon 34051, Republic of Korea\\
$^{2}$ \quad Department of Physics, KAIST, Daejeon 34141, Republic of Korea\\
$^{3}$ \quad RIKEN Center for Quantum Computing (RQC), Wako, Saitama 351-0198, Japan\\
$^{4}$ \quad Department of Applied Physics, Graduate School of Engineering, The University of Tokyo, Bunkyo-ku, Tokyo 113-8656, Japan}

\corres{Correspondence: swyoun@ibs.re.kr}

\firstnote{These authors contributed equally to this work.}

\abstract{The cavity haloscope is among the most widely adopted experimental platforms designed to detect dark matter axions with its principle relying on the conversion of axions into microwave photons in the presence of a strong magnetic field. 
The Josephson parametric amplifier (JPA), known for its quantum-limited noise characteristics, has been incorporated in the detection system to capture the weakly interacting axion signals.
However, the performance of the JPA can be influenced by its environment, leading to potential unreliability of a predefined parameter set obtained in a specific laboratory setting. 
Furthermore, conducting a broadband search requires consecutive characterization of the amplifier across different tuning frequencies.
To ensure more reliable measurements, we utilize the Nelder-Mead technique as a numerical search method to dynamically determine the optimal operating conditions.
This heuristic search algorithm explores the multidimensional parameter space of the JPA, optimizing critical characteristics such as gain and noise temperature to maximize signal-to-noise ratios for a given experimental setup.
Our study presents a comprehensive analysis of the properties of a flux-driven JPA to demonstrate the effectiveness of the algorithm.
This approach contributes to ongoing efforts in axion dark matter research by offering an efficient method to enhance axion detection sensitivity through the optimized utilization of JPAs.}


\begin{document}

\tableofcontents

\section{Introduction}~\label{sec:Introduction}
Axions are pseudo-Goldstone bosons emerging from the Peccei-Quinn mechanism, proposed to solve the charge-parity symmetry problem in quantum chromodynamics (QCD)~\cite{PecceiQuinn:PRL:1977}.
In particular, a substantially large value of the spontaneous symmetry breaking scale gave rise to the ``invisible" axions, which are categorized into two classes of theoretical models within the QCD framework: Kim-Shifman-Vainshtein-Zakharov (KSVZ) \cite{Kim:PRL:1979, Shiftman:NPB:1980} and Dine-Fischler-Srednicki-Zhitnitsky (DFSZ)~\cite{Zhitnitsky:SJNP:1980, Dine:PLB:1981} models.
These particles are anticipated to be neutral, stable and exhibit minimal interaction with ordinary matter, presenting a viable candidate for dark matter~\cite{Preskill:PLB:1983, Abbott:PLB:1983, Dine:PLB:1983}, one of the most profound mysteries in modern science. 
This dual significance positioned the axion as a major focus in the fields of particle physics and cosmology. 

In search for these elusive particles, the cavity haloscope~\cite{Sikivie:PRL:1983} provides the most sensitive approach, especially in the microwave regime. 
In the presence of a strong magnetic field, the axions are expected to be converted into photons, with the signal resonantly enhanced by a cavity tuned to the axion Compton frequency.
The signal is amplified to a detectable level using a series of low-noise amplifiers.
Since the axion mass (equivalently frequency with $m_a\simeq\omega_a=2\pi\nu_a$) is unknown {\it a priori}, the experimental performance is determined by the speed of frequency scanning for a given sensitivity as~\cite{Kim:JCAP:2020}
\begin{equation}
    \frac{d\nu}{dt} \simeq \left(\frac{1}{\textrm{SNR}} \right)^{2} \left(\frac{P_{a\gamma\gamma}}{k_{B}T_{\textrm{sys}}}\right)^{2} \left(1 + \frac{Q_{a}}{Q_{c}}\right),
    \label{eq:scan_rate}
\end{equation}
where $P_{a\gamma\gamma}$ is the signal power of the axion-induced photons, SNR denotes the desired signal-to-noise ratio, and $Q_a$ and $Q_c$ stand for the quality factors of the axion and cavity, respectively.
The system noise is represented by the equivalent temperature $T_{\rm sys}$ according to the Johnson-Nyquist theorem with the Boltzmann constant $k_B$~\cite{Johnson:PR:1928,Nyquist:PR:1928}.

The system noise is decomposed into two parts: noise originating from the thermal background and noise introduced by the readout chain.
Thermal noise is dictated by the physical temperature of the system, which can be reduced to levels as low as 10\,mK by utilizing advanced cryogenic technologies such as dilution refrigerators.
In order to take Bose-Einstein statistics of thermal photons into account, the physical temperature was transformed to the effective temperature, defined as
\begin{equation}
    T_{\mathrm{eff}} = \frac{hf}{k_{B}}\left[\frac{1}{ (e^{hf/k_BT_{\rm phy} } - 1)} + \frac{1}{2} \right],
    \label{eq:T_eff}
\end{equation}
where $T_{\rm{phy}}$ represents the physical temperature of the test system and $h$ denotes the Planck constant.
A typical readout chain consists of a multistage amplifier and the total noise is primarily dominated by the noise figure of its first stage amplifier according to the Friis's formula~\cite{Friis:ProcIRE:1944}, requiring a low-noise amplifier is placed at the first stage of a receiver chain. 
Since an axion search experiment needs to scan a certain frequency range, the cavity and the amplifier are required to be tunable, so is the amplifier.
For these reasons, a tunable superconducting amplifier, such as Josephson Parametric Amplifier (JPA)~\cite{Yamamoto:APL:2008}, is suitable for the first amplification stage.
The minimum value of each noise component is governed by the uncertainty principle, quantified as half of the photon noise, resulting in the standard quantum limit of system noise for phase-insensitive detection, expressed as $T_{\rm SQL} = h\nu/k_{B}$.
Due to their high gain, low noise, and tunable operation, flux-driven JPAs are used in various applications requiring low-noise amplification.

JPAs exploit the nonlinear behavior of Josephson junctions to achieve parametric amplification.
In particular, a flux-driven JPA utilizes magnetic flux to control its resonance frequency and other operating parameters.
For axion search experiments, JPAs are typically characterized either across the entire tunable frequency range prior to the experiment, providing comprehensive information through intensive work, or dynamically at a specific frequency during the experiment, relying on a simpler \textit{in-situ} measurement method such as the two-point Y-factor technique~\cite{PhysRevD.97.092001} or the SNR improvement method~\cite{PhysRevLett.124.101303}.
The characteristics of JPAs, however, are highly influenced by environmental factors such as experimental temperature, RF interference, ambient magnetic field background, and others, which can vary throughout the course of the experiment.
Consequently, these approaches could be less reliable and/or more time-consuming.
With this in mind, we adopt the Nelder-Mead (NM) algorithm~\cite{NelderMead:CJ:1965} as a heuristic search method to find optimal operating conditions for the JPAs.
This algorithm directly finds the optimal working conditions of JPAs on a multi-dimensional parameter space for a given experimental environment, it enables \textit{in-situ} measurements for versatile characteristics of devices and substantially increases the reliability of measurements compared to the aforementioned approaches.
In this study, we provide a detailed description of the methodology for characterizing a flux-driven JPA using this algorithm.

\section{Materials and Methods}\label{sec:Materials_and_Methods}
\subsection{Flux-driven JPA}\label{sec:flux_driven_JPA}
\begin{figure}
\centering
\includegraphics[width=.75\linewidth]{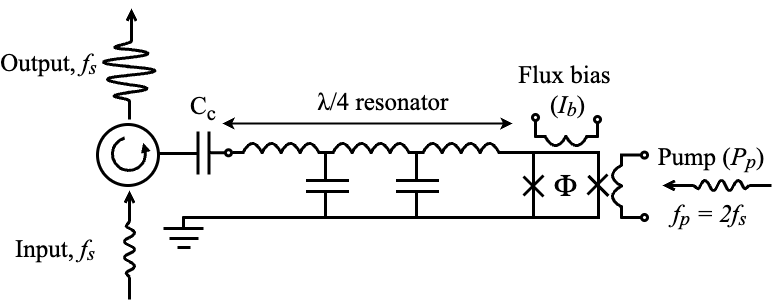}
\caption{Equivalent circuit diagram of a flux-driven JPA consisting of a $\lambda/4$ coplanar waveguide resonator terminated by a SQUID defined by the Josephson junctions denoted as $\times$.
The inductance of the SQUID is set by the magnetic flux bias which is induced by the bias current ($I_b$) and modulated by the pump signal ($P_p$) at twice the frequency of the signal to parametrically amplify the desired signal. 
Since the signal input and output share the same port, a circulator is required to regulate the direction.}
\label{fig:jpa_diagram}
\end{figure}

The basic design and operation principles of flux-driven JPAs are described in Ref.~\cite{Kutlu_2021}.
The flux-driven JPA consists of a Superconducting Quantum Interference Device (SQUID) at the end of a quarter-wave ($\lambda/4$) coplanar waveguide resonator.
This resonator, capacitively coupled to a transmission line, serves as a common path for both the input and output signal.
Figure~\ref{fig:jpa_diagram} shows the equivalent circuit diagram of the JPA, including the directions of the input and output signals with respect to the amplifier.

The flux-driven JPA utilizes SQUIDs to detect flux changes induced by modulated Josephson junctions, enabling parametric amplification of weak signals.
The SQUID acts as an inductor in a resonant circuit whose inductance changes non-linearly with respect to the magnetic flux passing through the SQUID loop.
This magnetic flux is generated by a DC bias current ($I_b$) applied to a superconducting coil adjacent to the loop.
Parametric amplification of signals is achieved by modulating the SQUID inductance through an external pump signal through a line inductively coupled to it.
A three-wave mixing process involves pump ($p$), signal ($s$), and idler ($i$) waves with a frequency relationship of $f_{p} = f_{s} + f_{i}$.
Since the JPA receives the signal and reflects it back through the same port, a circulator is utilized to separate the incoming and outgoing signals.
The operating frequency can be tuned over a certain range by adjusting the DC bias current, thereby varying the average inductance of the SQUID.
The optimal operation also requires a fine-tuning of the power of the pump signal to maximize the amplification of the desired signal while minimizing undesirable effects such as distortion or noise.
The JPA is protected from external magnetic fields by a multi-layer shield consisting of aluminum, Cryophy and NbTi alloy layers from innermost to outermost, respectively~\cite{Sergey:LT29:2023}. 
The corresponding shielding coefficient was estimated to be between $10^{5}$ and $10^{6}$, which is sufficiently high for reliable JPA operation in environments with residual magnetic fields up to 0.1\,T.

\subsection{Nelder-Mead Algorithm}~\label{sec:NMalgorithm}
The Nelder-Mead algorithm provides a numerical approach designed to locate stationary points (local minima or maxima) of objective functions in a multidimensional parameter space.
Serving as a heuristic search technique, it is particularly well-suited for addressing nonlinear optimization problems, especially when calculating derivatives is either impractical or undesirable.
The algorithm converges to a stationary point, but not necessarily to the global optimum.
This algorithm utilizes the concept of a simplex, a geometric shape comprising $n+1$ points in an $n$-dimensional space. 
Throughout the optimization process, this simplex undergoes various transformations, including reflection, expansion, contraction, and shrinkage. 
By iteratively adjusting the simplex, the algorithm explores the search space to approximately locate the minimum/maximum value of the objective function.
The algorithm continues to iterate until it meets a specified termination criterion, either reaching a maximum number of iterations or fulfilling a predetermined level of requirement.
Through this iterative process, the algorithm returns the best-found, albeit local, solution, i.e., the vertex with the lowest function value.

\subsection{Measurement setup and calibration}~\label{sec:Meas_and_cal}
\begin{figure}
\centering
\includegraphics[width=.65\linewidth]{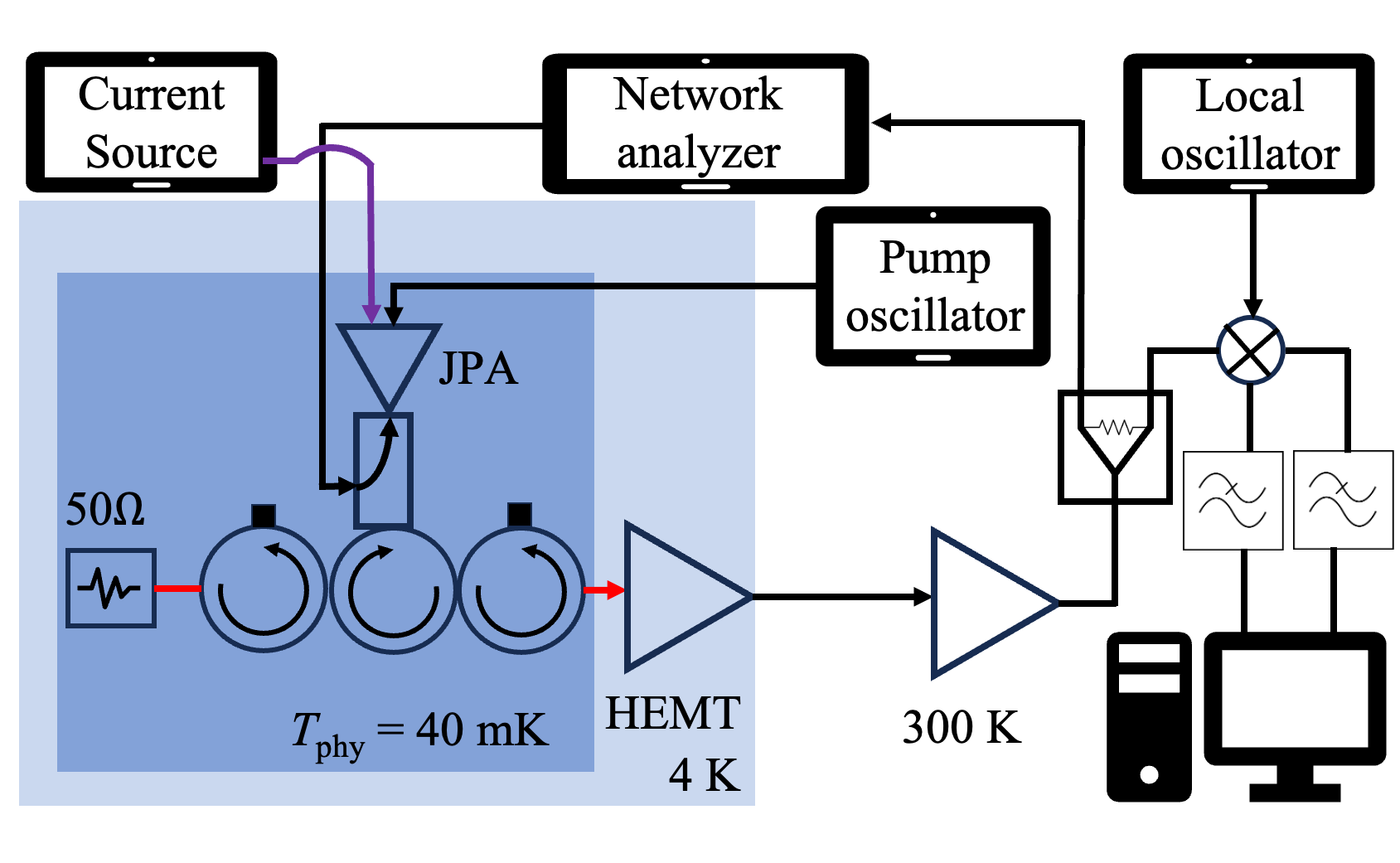}
\caption{Schematic of the JPA characterization setup.
The major RF components include the 50-$\Omega$ NS, a set of circulator/isolators, a directional coupler, and a series of HEMTs.
The operation of the JPA requires a flux bias and a pump signal supplied by a current source and an oscillator, respectively.
The red and purple lines indicate the RF superconducting cables and the DC bias line, respectively.}
\label{fig:exp_schematic}
\end{figure}

The overall measurement setup for the JPA characterization is schematized in Fig.~\ref{fig:exp_schematic}.
The dilution refrigerator (DR) has a capability of cooling the major components, including JPA, circulator/isolators, and the noise source (NS), attached to the mixing-chamber plate (MXC) to temperatures below 30\,mK.
The RF devices were strongly anchored to the MXC for thermalization, while the NS was weakly connected to the MXC for thermal isolation.
For stable operation of the JPA, the MXC temperature was maintained at 40\,mK using a proportional-integral-differential controller.
A series of a circulator and isolators were configured to i) provide a dual path for noise signal when characterizing the JPA and for axion signal when performing an axion search experiment and ii) reduce interference between the RF components.
The pump signal from a signal generator was directly fed into the JPA and the bias current was supplied using a high-precision current source through a superconducting DC line.
A directional coupler was introduced to measure the JPA gain through transmission measurements using a vector network analyzer (VNA).
An attenuation of about 110\,dB was placed between the VNA and JPA in order to prevent JPA saturation caused by the VNA output power, which will be discussed in Section~\ref{sec:JPA_satruation}.
The noise source, a 50-$\Omega$ terminator with a heater attached, provided the noise signal for calibration via a superconducting RF line.
The circuit was extended to the 4-K stage where a pair of high electron mobility transistors (HEMTs) were connected via another superconducting transmission line.
A 20-dB attenuator was placed between the HEMT amplifiers to diminish the interference.
The signals were further amplified at room temperature, and underwent a series of processes including down-conversion to an intermediate frequency, digitization, Fourier transformation and storage on tape.

\begin{figure}
\centering
\includegraphics[width=.6\linewidth]{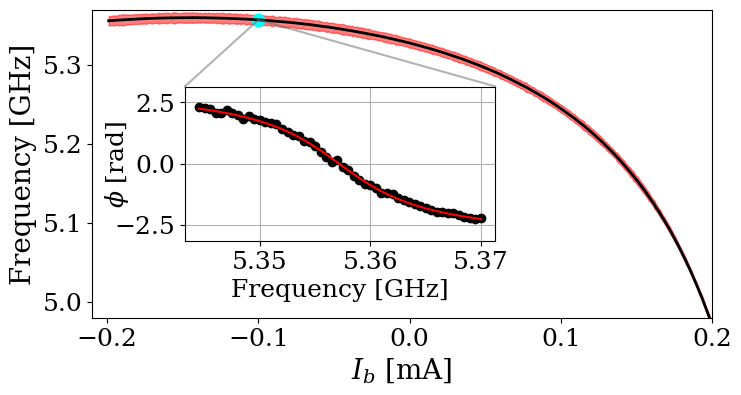}
\caption{Passive resonant frequency of the JPA as a function of flux bias ($I_b$), with the red band representing the bandwidth.
The inset provides an example of the phase response fitted with the arc-tangent transfer function, given in Equation~\ref{eq: JPA_phase}, at $I_b=-0.1$\,mA.
}
\label{fig:flux_sweep}
\end{figure}

Prior to characterizing the JPA, the gain and noise temperature of the receiver chain, excluding the JPA, were assessed using the Y-factor method.
The output noise power levels were measured at various temperatures of the noise source ranging from 100\,mK to 500\,mK at 100-mK intervals.
The noise power exhibited good linearity with respect to the effective temperature (Equation~\ref{eq:T_eff}), and a linear fit resulted in a gain of 74--75 dB and a noise temperature of 3.0--3.5 K over a broad frequency range spanning from 4.9\,GHz to 5.4\,GHz.
The tunable range of the JPA was determined by observing passive resonances while sweeping the flux bias with the pump signal turned off.
To find the resonant frequency for each flux bias, the phase response obtained using the VNA was fitted near resonance with an arc-tangent transfer function of
\begin{equation}
    \phi = A\arctan\left(\frac{f-f_r}{\Delta f}\right),
    \label{eq: JPA_phase}
\end{equation}
where $A$, $f_r$, and $\Delta f$ are the amplitude, resonant frequency and bandwidth, respectively. 
A set of $f_r$ and $\Delta f$ values obtained over multiple flux quanta constructed a flux sweep curve.
A segment of the curve corresponding to approximately half a flux quantum is shown in Figure~\ref{fig:flux_sweep}, from which the operating resonant frequency range was determined to be approximately 5.00\,GHz to 5.35\,GHz.

\subsection{JPA saturation}~\label{sec:JPA_satruation}
\begin{figure}
\centering
\includegraphics[width=\linewidth]{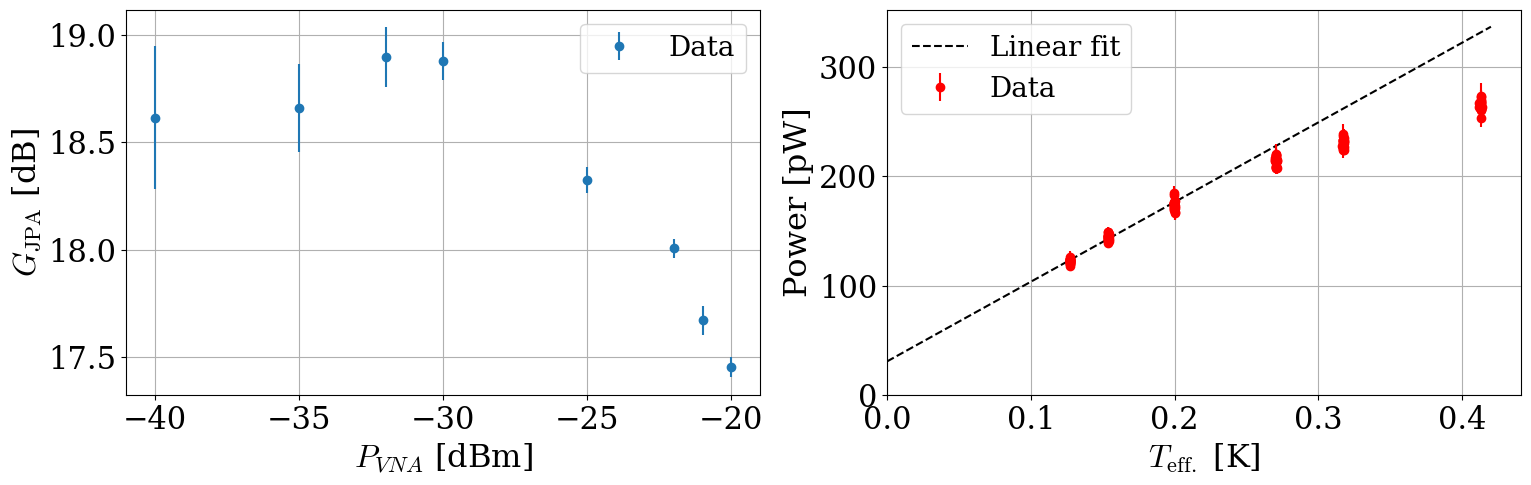}
\caption{Exploration of the JPA saturation by analyzing the JPA gain response to VNA power (left) and the dependence of noise power on the effective temperature of the noise source (right), with respect to varying input power levels.
The measurements were conducted at a frequency 200\,kHz offset from the half pump frequency of 5.225\,GHz.
}
\label{fig:power saturation}
\end{figure}

JPAs rely on the nonlinearity in the Josephson junction, meaning the gain of the JPA changes with variations in the amplitude of input signals.
Exceeding certain input power levels leads to junction saturation, which reduces the JPA gain and distorts the output signal, thus limiting dynamic range and linearity. 
Therefore, maintaining the input power at appropriate levels is crucial for optimizing JPA performance and preventing saturation.
This requires a precise understanding of the saturation power to ensure linear operation within the optimal range.
Such power saturation issues may arise in certain measurement processes, e.g., JPA gain measurement via transmission measurements using a VNA and noise figure evaluation through the Y-factor method involving a noise source.
The saturation power of our JPA was independently estimated using the following methods.
The JPA gain, determined as the ratio between output signals in the JPA on and off states, was calculated across a range of VNA output power levels spanning from $-$40\,dBm to $-$20\,dBm.
In the Y-factor measurement, noise power was acquired at various physical temperatures of the noise source, ranging from 40\,mK to 500\,mK.
These measurements were performed at a half pump frequency of 5.23\,GHz, and the results are shown in Figure~\ref{fig:power saturation}.
A gradual decrease in gain is evident at high VNA output powers, and the non-linearity becomes apparent at high noise temperatures.
Based on these findings, the acceptable maximum VNA power and NS temperature for reliable JPA operation were found to be about $-$30\,dBm and 200\,mK, respectively.
These numbers consistently indicate that the saturation power of the JPA is approximately $-$140\,dBm, accounting for all attenuation from upstream components in our measurement setup and considering the typical JPA bandwidth of around 1\,MHz.
The Y-factor technique using the data points up to 200\,mK resulted in a JPA gain and system noise temperature of $G_{\rm JPA} = 18.7\pm0.3$\,dB and $T_{\rm sys}^{\rm on} = 336.0 \pm 5.5$\,mK, respectively. 
The obtained gain coincides with that acquired using the VNA, validating the saturation power measurements.

\subsection{Noise visibility ratio}~\label{sec:NVR}
\begin{figure}
\centering
\includegraphics[width=.9\linewidth]{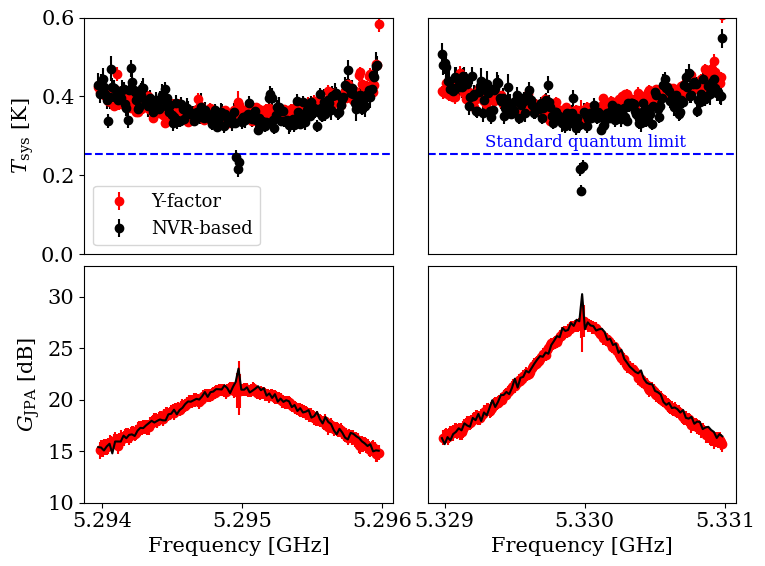}
\caption{Comparison of noise temperature (top row) and gain (bottom row) measurements of the JPA using both NVR-based (black) and Y-factor (red) methods under two different operating conditions. 
Deviations in data points at the half pump frequency are attributed to signa-idler degeneracy.
Both measurement techniques exhibit a mutual agreement within 2\%.
}
\label{fig:JPA_Y_fact}
\end{figure}

One of the key characteristics of JPAs for many applications is their noise performance, which can be assessed using the noise visibility ratio (NVR)~\cite{Friis:ProcIRE:1944, PhysRevX.5.041020}.
It is defined as the ratio of the excess noise observed in the power spectrum when the device is activated versus deactivated:
\begin{equation}
    {\rm NVR} = \frac{P^{\rm{on}}}{P^{\rm{off}}} = \frac{G_{\rm{JPA}} T_{\rm{sys}}^{\rm on}}{T_{\rm{sys}}^{\rm off}},
\end{equation}
where $P^{\rm{on, off}}$ denotes the power spectrum level obtained with the JPA turned on and off, respectively. 
The system noise temperature with the JPA off, $T_{\rm{sys}}^{\rm off}$, is determined beforehand using the Y-factor method as discussed in Section~\ref{sec:Meas_and_cal}. Additionally, the gain of the JPA, $G_{\text{JPA}}$, is measured using a VNA. 
This technique facilitates a time-efficient {\it in-situ} measurement of noise temperature, offering an alternative approach to more challenging direct methods, such as the Y-factor method.
The methodological validation was made by comparing it with the Y-factor method. 
The JPA gain and system noise temperature were independently measured at certain $I_b$ and $P_p$ values and subsequently compared.
For the Y-factor measurement, the physical temperature of the noise source was varied from 40\,mK to 200\,mK, a range chosen to avoid JPA saturation (see Section~\ref{sec:JPA_satruation}).
Figure~\ref{fig:JPA_Y_fact} shows the measurement results at two different frequencies, illustrating that the two approaches are in agreement within a margin of 2\%.
The data points were fitted with a quadratic polynomial to obtain the noise temperature and with a Lorentzian function to obtain the JPA gain, both evaluated 200\,kHz away from half the pump frequency in order to operate the JPA in the non-degenerate mode.

\section{Results}\label{sec:Result}
\subsection{JPA parameterization}\label{sec:JPA_parameterization}
\begin{figure}
\centering
\includegraphics[width=\linewidth]{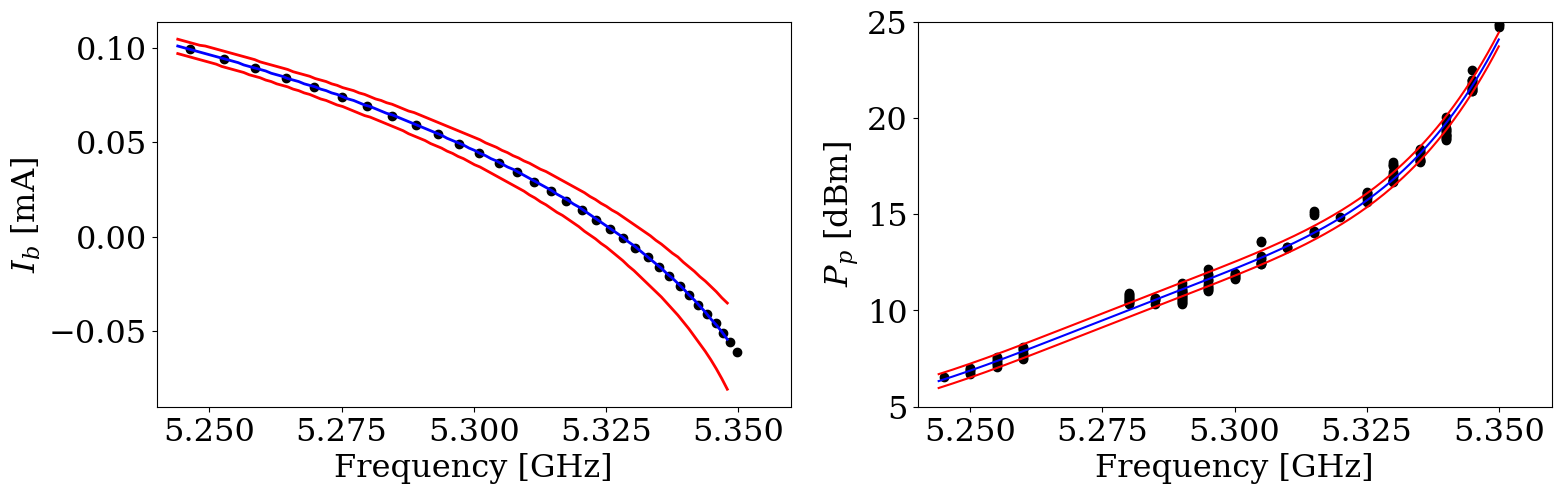}
\caption{
Parameterization of the key parameters for optimal operation of the JPA as a function of frequency: bias current (left) and pump power (right).
The passive resonance curve was fitted using an effective function described in the text, while the pump power profile was fitted using a 5th-order polynomial.
The blue lines represent the central values of the fit and the pairs of red lines indicates 10\% bandwidth and averaged RMS value, respectively.
}
\label{fig:parameterization}
\end{figure}

The operation of JPAs typically relies on adjusting two main parameters: i) bias current to tune the JPA frequency, and ii) pump power to achieve parametric amplification of signals.
These two parameters are crucial for controlling the behavior and performance of JPAs, allowing for optimization of key characteristics such as gain, bandwidth, and noise figure.
In order to establish the initial simplex for the NM algorithm at a given resonant frequency, these parameters need to be pre-measured and parameterized as a function of frequency.
The bias current is swept to obtain a profile of the JPA passive resonance frequency as a function of bias current, as discussed in Section~\ref{sec:Meas_and_cal}.
This process involves varying the bias current to observe changes in JPA resonance, thereby mapping out the tunable range of the JPA and identifying the bias current values corresponding to the desired resonant frequencies.
The pump power was also swept to acquire a set of values that yielded a JPA gain of $20\pm5$\,dB using the VNA.
This step entails systematically adjusting the pump power while monitoring the resulting gain to identify the optimal pump power for each frequency. 
These two parameters were parameterized by fitting the acquired data points as a function of frequency, as shown in Figure~\ref{fig:parameterization}.
The fitting function for the passive resonance curve was adopted from Ref.~\cite{PhysRevB.74.224506}, which models the response of the effective SQUID inductance to the bias current through an equivalent LC circuit.
The pump power profile was fitted using a polynomial function of degree 5.
The red lines correspond to a 10\% bandwidth of JPA resonance for the former and an averaged RMS value of 0.4\,dBm from the fitted value for the latter.
Our study is limited to the frequency range between 5.24 and 5.35 GHz, over which the JPA was utilized in an axion search experiment~\cite{kim2023experimental}.
These pre-measured parameter sets serve as initial test points of the simplex for the NM algorithm, enabling it to iteratively refine the solution towards the desired operation conditions.

\subsection{JPA characterization}~\label{sec:JPA_characterization}
The NM algorithm was integrated with the NVR method to find the optimal parameter values for achieving high-performance JPA operation.
In our application, the parameter space consists of the bias current ($I_{b}$) and pump power ($P_{p}$), and thus a triangle simplex was chosen.
For a given resonant frequency, the algorithm constructs the initial simplex with the test points based on values derived from the parameterization process described in Section~\ref{sec:JPA_parameterization}: one consisting of the fitted values ($I_{b}^{0},P_{p}^{0}$) and the other two of the values offset by the amount of the 10\% fitting uncertainty and deviation ($I_{b}^{0}-\Delta I_{b}^{0}, P_{p}^{0}- \Delta P_{p}^{0}$ and $I_{b}^{0}+\Delta I_{b}^{0}, P_{p}^{0} - \Delta P_{p}^{0}$).
With these initial simplex points, the NM algorithm coefficients for reflection, expansion, contraction, and shrinkage are set to $1, 2, 0.5,$ and $0.5$, respectively. 
The algorithm evaluates the performance (in this case, noise temperature) at each test point, extrapolates to find a new point using one of the transformations, and replaces the worst-performing test point with this new one, potentially improving overall performance.

\begin{figure}
\centering
\includegraphics[width=.65\linewidth]{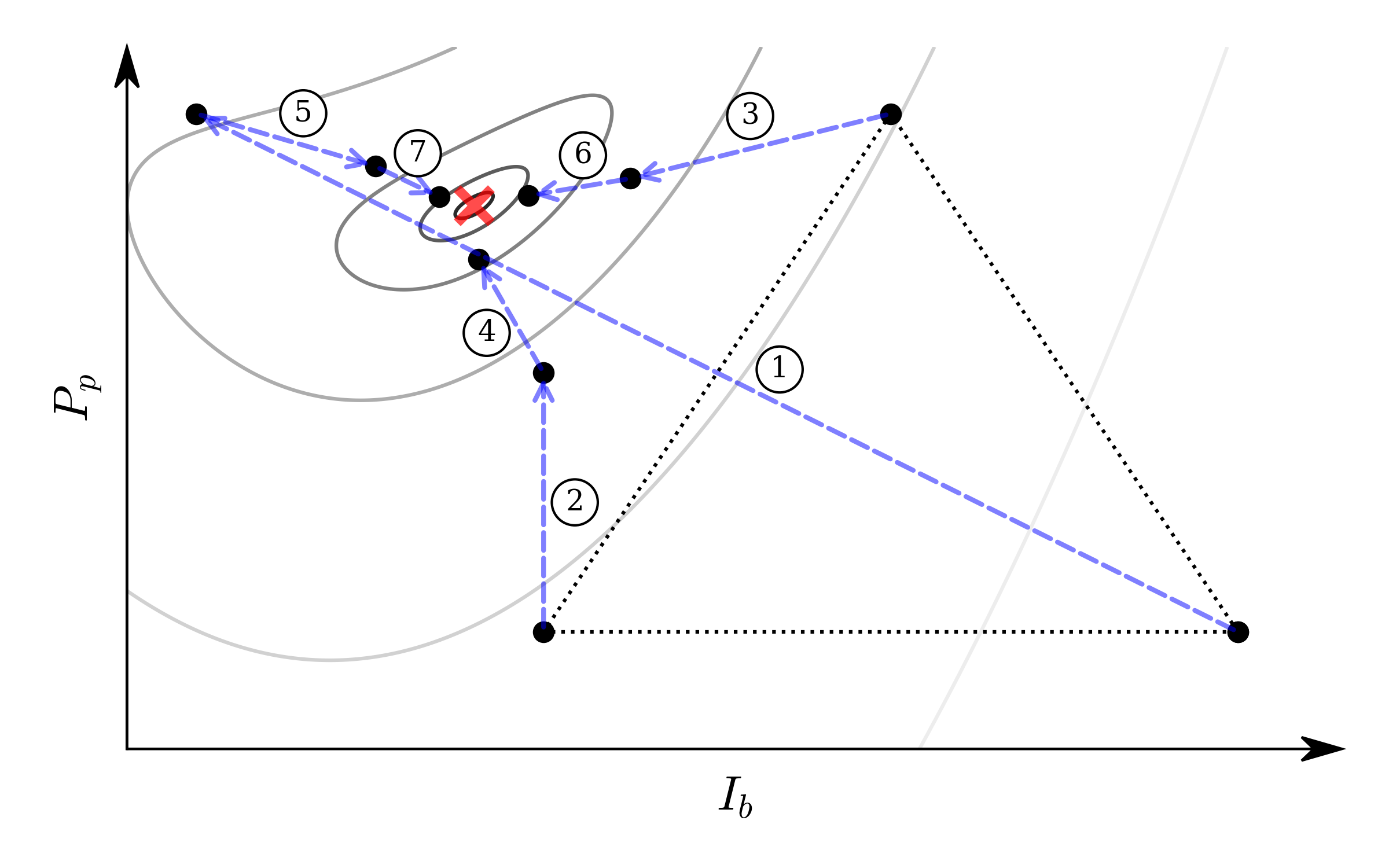}
\caption{Visualization of the Nelder-Mead method to search for the minimum noise temperature in the two-dimensional space ($I_b$, $P_p$).
The contour lines depict different noise temperature levels, with the true minimum value indicated by the red cross.
The initial simplex is represented by black dotted lines, while the iteration process is illustrated through a sequence of transformations denoted by blue dashed lines with numerical labels.
}
\label{fig:initial_simplex}
\end{figure}

\begin{figure}
\centering
\includegraphics[width=\textwidth]{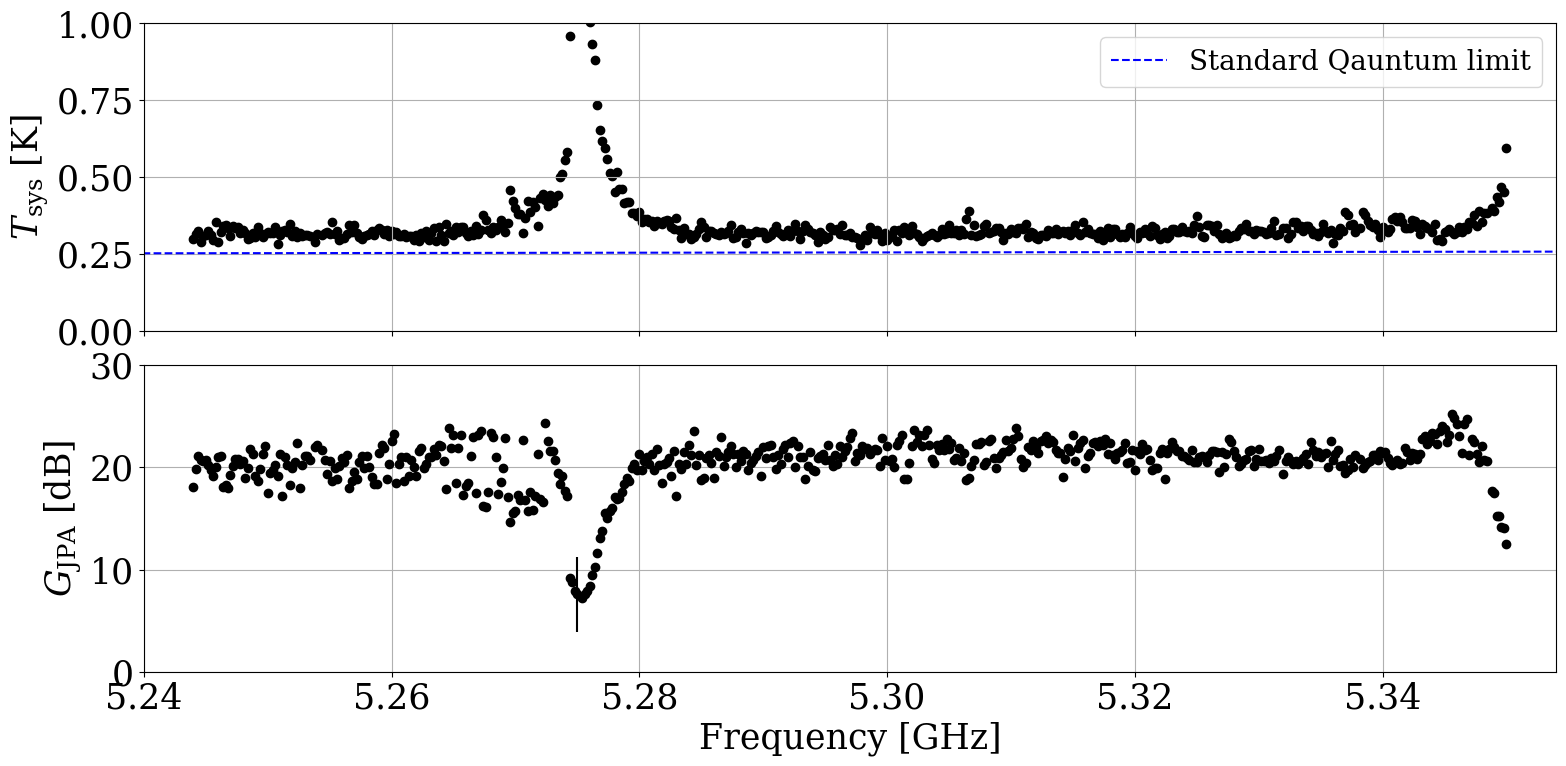}
\caption{System noise temperature (top) and JPA gain (bottom) found by the NM algorithm as a function of frequency. 
The blue dashed line represents the standard quantum limit.
The anomalous behavior observed around 5.275\,GHz and 5.35\,GHz is presumably attributed to impedance mismatch and maximum resonant frequency, respectively.}
\label{fig:NM_result}
\end{figure}

\begin{figure}
\centering
\includegraphics[width=0.6\linewidth]{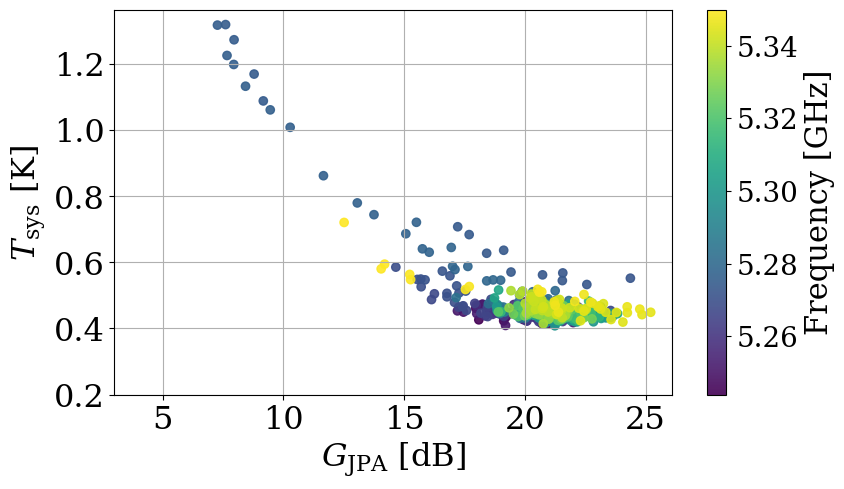}
\caption{Correlation between the system noise temperature and JPA gain derived from the data presented in Figure~\ref{fig:NM_result}. 
}
\label{fig:NM_result2}
\end{figure}

In each iteration, the new test point was evaluated in terms of power saturation of the JPA using the VNA.
An increase of more than 10\% in JPA gain for VNA power 3-dB weaker than nominal was considered indicative of saturation. 
In such cases, a new test point was set with a pump power reduced by 0.01 dB, and the evaluation process was repeated.
Once the requirement was met, the noise temperature was measured using the NVR technique.
The number of iterations was set to 10, a value determined empirically by observing convergence of the noise temperature to its minimum.
The iteration process typically takes 2 to 3 minutes to complete, depending on the initial values of the seed and the occurrence of power saturation.
The NM algorithm provides an \textit{in-situ} method to characterize the JPA, thereby substantially increasing the reliability of measurements compared to predetermined parameter sets obtained during the commissioning phase of an experiment.
The entire process was repeated across a frequency range from 5.244--5.350\,GHz, with a tuning step of 200\,kHz selected to be approximately half the bandwidth of the cavity used for axion search.
The resulting system noise temperature and the corresponding JPA gain are presented in Figure~\ref{fig:NM_result} as a function of frequency.
Taking into account $T_{\rm SQL}=254$\,mK at 5.3\,GHz, the system noise corresponds to approximately 1.5 noise photons and the JPA gain remains around 20\,dB.
The uncertainties in measurements were determined from the fitting errors, which are 3.3\,mK for the system noise temperature and 0.04\,dB for the JPA gain.
It is also noteworthy that the NM algorithm found the solutions even around abnormal regions, e.g., at around 5.275\,GHz.
It was observed that slight variations in configuration, such as using different circulators or adjusting line lengths, resulted in shifts of these regions to different frequencies. 
This suggests that such regions are presumably caused by impedance mismatch.
Further investigation will be necessary to gain a clear understanding of the underlying cause.
Finally, in Figure~\ref{fig:NM_result2}, the correlation between $T_{\rm{sys}}^{\rm on}$ and $G_{\rm{JPA}}$ is shown, highlighting an inverse relationship between these two parameters.
This confirms the Friis's formula that the noise contribution from downstream components increases as the gain of the first stage amplifier decreases.

\section{Discussion and conclusion}~\label{sec:discussion}
In summary, JPAs serve as a key device essential for achieving low-noise amplification in a variety of fields, especially highly sensitive applications such as the search for axion dark matter.
Flux-driven JPAs are especially advantageous for exploring a broad frequency range, although they require careful \textit{in-situ} characterization while ensuring reliability in versatile environments. 
This study leverages the Nelder-Mead algorithm to significantly improve the efficiency and reliability of JPA characterization.
By adopting such a systematic approach, JPAs can be leveraged to support groundbreaking research efforts, including the exploration of QCD axion models.
The successful characterization of a flux-driven JPA operating around 5.3\,GHz demonstrates the effectiveness of this method, validating its potential to enhance current and future axion dark matter searches.
This method holds significant promise to play a pivotal role in advancing the field, providing a powerful tool for uncovering deep insights into the nature of dark matter and related phenomena.

\vspace{6pt} 

\authorcontributions{Conceptualization, S.Y.; methodology, Y.K. and J.J.; software, Y.K. and J.J.; validation, Y.K., J.J., S.Y. and S.B.; formal analysis, Y.K. and J.J.; investigation, Y.K. and J.J.; resources, A.F.L., Y.N. and S.U.; data curation, Y.K., J.J. and S.Y.; writing---original draft preparation, J.K.; writing---review and editing, S.Y.; visualization, Y.K., J.J. and S.Y.; supervision, S.Y.; project administration, S.Y. and Y.S.; funding acquisition, Y.S. All authors have read and agreed to the published version of the manuscript.}

\funding{This research was funded by the Institute for Basic Science (IBS-R017-D1) and JSPS KAKENHI (JP22H04937).}



\dataavailability{Data sharing is available by emailing the corresponding author.} 


\conflictsofinterest{The authors declare no conflicts of interest.} 

\abbreviations{Abbreviations}{
The following abbreviations are used in this manuscript:\\

\noindent 
\begin{tabular}{@{}ll}
QCD & Quantum chromodynamics\\
JPA & Josephson parametric amplifier\\
DR & Dilution refrigerator\\
MXC & Mixing-chamber plate\\
VNA & Vector network analyzer\\
HEMT & High electron mobility transistor\\
NM & Nelder-Mead \\
NVR & Noise Visibility Ratio \\
\end{tabular}
}

\begin{adjustwidth}{-\extralength}{0cm}
\reftitle{References}
\bibliography{main}
\PublishersNote{}
\end{adjustwidth}

\end{document}